\documentclass[twoside,english,sort&compress]{iopart}
\usepackage[T1]{fontenc}
\usepackage[latin9]{inputenc}
\usepackage{geometry}
\usepackage{color}
\usepackage{xcolor}
\usepackage{float}
\usepackage{graphicx}
\usepackage{esint}
\usepackage[numbers]{natbib}

\makeatletter
\usepackage{iopams}
\usepackage{setstack}

\newcommand{\eqref}[1]{(\ref{#1})}

\makeatother

\usepackage{babel}
\begin{document}
\title{Experimental reconstruction of spatial Schmidt modes for a wide-field SU(1,1) interferometer}
\author{Gaetano Frascella, Roman V. Zakharov, Olga V. Tikhonova and Maria
V. Chekhova}
\address{1 Max-Planck Institute for the Science of Light, Staudtstr. 2, Erlangen
D-91058, Germany}
\address{2 University of Erlangen-Nuremberg, Staudtstr. 7/B2, 91058 Erlangen,
Germany}
\address{3 Physics Department, Moscow State University, Leninskiye Gory 1-2,
Moscow 119991, Russia}
\address{4 Skobeltsyn Institute of Nuclear Physics, Lomonosov Moscow State
University, Moscow 119234, Russia}
\ead{{\large{}gaetano.frascella@mpl.mpg.de}}
\begin{abstract}
We study the spatial mode content at the output of a wide-field ${\rm SU(1,1)}$ interferometer, i.e. a nonlinear interferometer comprising two coherently-pumped spatially-multimode optical parametric amplifiers placed in sequence with a focusing element in between. This device is expected to provide a phase sensitivity below the shot-noise limit for multiple modes over a broad angular range. To reconstruct the spatial modes and their weights, we implement a simple method based on the acquisition of only intensity distributions. The eigenmode decomposition of the field is obtained through the measurement of the covariance of intensities at different spatial points. We investigate both the radial and azimuthal (orbital angular momentum) modes and show that their total number is large enough to enable applications of the interferometer in spatially-resolved phase measurements.
\end{abstract}

\maketitle

\section{Introduction}

A lot of interest is currently attracted to the so-called ${\rm SU(1,1)}$ interferometer \citep{Yurke:86,Hudelist:14,Manceau:17PRL,Anderson:17}. This is a sequence of two coherently pumped optical parametric amplifiers, based on four-wave mixing (FWM)~\citep{Hudelist:14,Anderson:17} or parametric down-conversion (PDC)~\citep{Manceau:17PRL}. The second amplifier amplifies or de-amplifies the radiation from the first one depending on the phase shift in between, and the interferometer provides sub-shot-noise sensitivity to this phase. Recently a wide-angle version of such an interferometer has been realized in experiment~\citep{Frascella:19}, aimed at various applications, among them testing the spatially-distributed phase shifts. In particular, sub-shot-noise sensitivity to the orbital angular momentum (OAM) is expected~\citep{Beltran:17,Liu:18}. All these applications require a large number of spatial modes covered by the interferometer.

Experimentally quantifying the spatial mode content of a multimode radiation can be a challenging task. In general, the measurement of the spatial intensity distribution is not sufficient because of the
overlapping of different modes at each point. For this reason, more advanced techniques need to be used to access the information about the radiation and its orthogonal spatial components.

In particular, the measurement of the OAM spectrum has raised special interest since its discovery \citep{Allen:92}. The promise of increased capacity for quantum communication due to
the discrete high-dimensional degree of freedom of OAM (unlike polarization) can be only fulfilled if an efficient detection of the mode spectrum is implemented \citep{Bozinovic:13,Larocque:17,Reichert:18,Pan:19}. Different modes are separated by exploiting the periodical dependence
of the phase of the OAM field on the azimuthal angle, either directly or with interferometric setups.

One direct strategy is to flatten this dependence with a spatial light modulator (SLM) for one OAM mode at a time and measure the coupling of the diffracted beam into a single-mode fiber \citep{Mair:01,Giovannini:12}. This strategy provides the distribution of the mode weights, but with
some limitations \citep{Qassim:14}. Alternatively, the OAM charge of a beam can be converted into the transverse position with two phase elements~\citep{Mirhosseini:13,Malik:14,Larocque:17}, which can be replaced by SLMs~\citep{Berkhout:10}. In this case, different OAM modes focus through a lens into separate spots, whose intensity can be simultaneously measured. It is even
possible to profit from the conservation of OAM in frequency upconversion to detect the modes \citep{Sephton:19}. 

The information about the OAM spectrum of a beam can be also indirectly determined by Fourier transforming the mutual angular coherence function, i.e. the first-order angular correlation function (CF) \citep{Jha:11}. Experimentally, one extracts the CF from the measurement of the visibility
from the interference of the beam under study and its rotated replica~\citep{Peeters:07,Pires:10}. However, such experiments rely on the fragile stability of interferometers and coincidence techniques over time. A novel version of this technique demonstrates the measurement of
the CF through single-shot acquisition of interferograms \citep{Kulkarni:17,Kulkarni:18},
therefore eliminating stability problems. Similarly, this issue does not play a role if the interference is observed between two different azimuthal parts of the same beam, like in a double-slit experiment
\citep{Malik:12}.

All these techniques involve a special detection setup with several optical elements. In this work, we use a technique for reconstructing the shapes and weights of eigenmodes by only acquiring a sufficient number of 2D intensity distributions. Our method  is valid for any radiation with Gaussian
statistics including thermal light and bright twin beams generated through high-gain PDC or FWM. The technique relies on the Siegert relation between the first- and second-order CFs. Indeed, we reconstruct the eigenmodes starting from the measurement of the second-order CF, i.e.
from the spatial correlations of intensity fluctuations. The main advantage for such a method is that the detection stage contains only a camera \citep{Defienne:18}. Moreover, under reasonable conditions
on the symmetry of the intensity 2D distribution, one can reconstruct independently the modes in the two degrees of freedom involved. For example, in the case of cylindrical symmetry one can reconstruct OAM
and radial modes. A similar method can be applied to the time/frequency domain if the detection device is fast enough to follow the intensity fluctuations in time~\citep{Finger:17}. The only limitation of our method is that the phases of radial modes can not be recovered.

Here we apply this method to test a wide-angle SU(1,1) interferometer~\citep{Frascella:19} consisting of two quadratically nonlinear crystals with a focusing element in between. Both crystals generate high-gain PDC, and the second crystal amplifies or de-amplifies the radiation generated in the first one, depending on the phase introduced into the pump beam. At the output, the radiation is multimode both spatially and temporally, and each mode is populated by bright squeezed vacuum (BSV). If signal and idler beams can be distinguished (for instance, by considering non-degenerate frequencies),  the BSV manifests twin-beam photon-number correlations~\citep{Perez:14,Spasibko:12}. Each beam taken separately manifests thermal photon statistics.

Below, we present a detailed study of the angular (radial and azimuthal) eigenmodes and their weights for the output radiation of the wide-angle SU(1,1) interferometer. The OAM modes are related to the azimuthal angle but their spectrum also depends on the radial angle selected. We focus our attention on the dependence of the OAM content on the radial angle, but also on the spectrum averaged
over all possible radial angles. We show, in particular, that for larger radial angles there is a richer spectrum of OAM modes.

\section{Covariance of intensities and the Schmidt decomposition\label{sec:Schmidt-decomposition}}

We derive in this Section the general result that links the measurement
of the correlations of the intensity fluctuations of any Gaussian field
with its eigenmode decomposition. For a Gaussian field, all high-order CFs can be expressed as a function
of the first-order one. The normally-ordered first- and second-order spatial CFs in the far field
are defined as
\begin{equation}
G^{(1)}(\vec{q},\vec{q'})=\left\langle {E}^{(-)}(\vec{q}){E}^{(+)}(\vec{q'})\right\rangle, \label{eq:G1}
\end{equation}
\begin{equation}
G^{(2)}(\vec{q},\vec{q'})=\left\langle {E}^{(-)}(\vec{q}){E}^{(-)}(\vec{q'}){E}^{(+)}(\vec{q}){E}^{(+)}(\vec{q'})\right\rangle, \label{eq:G2}
\end{equation}
with ${E}^{(+/-)}$ being the positive- and negative-frequency fields, while
$\vec{q}$ and $\vec{q'}$ the transverse wavevectors of two points
in the far field. The link between the two CFs can be found by using
the moment theorem for complex Gaussian processes \citep{MandelWolf},
\begin{equation}
\begin{array}{c}
\left\langle {E}^{(-)}(\vec{q}){E}^{(-)}(\vec{q'}){E}^{(+)}(\vec{q}){E}^{(+)}(\vec{q'})\right\rangle =\\
=\left\langle {E}^{(-)}(\vec{q}){E}^{(+)}(\vec{q})\right\rangle \left\langle {E}^{(-)}(\vec{q'}){E}^{(+)}(\vec{q'})\right\rangle +\\ \left\langle {E}^{(-)}(\vec{q}){E}^{(+)}(\vec{q'})\right\rangle \left\langle {E}^{(-)}(\vec{q'}){E}^{(+)}(\vec{q})\right\rangle .
\end{array}\label{eq:gaussth}
\end{equation}
Indeed, the substitution of Eq.~(\ref{eq:gaussth}) into Eq.~(\ref{eq:G2})
and the definition in Eq.~(\ref{eq:G1}) leads to the well-known Siegert relation 
\begin{equation}
G^{(2)}(\vec{q},\vec{q'})=\langle {I}(\vec{q})\rangle \langle {I}(\vec{q'})\rangle +\left|G^{(1)}(\vec{q},\vec{q'})\right|^{2},
\end{equation}
{where ${I}(\vec{q})={E}^{(+)}(\vec{q}){E}^{(-)}(\vec{q})$ is the intensity.}

In Sec. \ref{sec:SU(1,1)} we measure a related quantity, namely the covariance $\rm{Cov}(\vec{q},\vec{q'})=G^{(2)}(\vec{q},\vec{q'})-\langle {I}(\vec{q})\rangle \langle {I}(\vec{q'})\rangle $.
The square root of this quantity is then directly proportional to
the absolute value of the first-order CF and this result links the covariance with the coherent modes of the system. Indeed, according to the classical coherence theory, the first-order CF for a multimode
beam can be written in the form 
\begin{equation}
G^{(1)}(\vec{q},\vec{q'})=\sum_{m,n}\tilde{\lambda}_{mn}\tilde{u}_{mn}^{*}(\vec{q})\tilde{u}_{mn}(\vec{q'}),\label{eq:cohmodes}
\end{equation}
where $\tilde{u}_{mn}$ are the eigenfunctions and $\tilde{\lambda}_{mn}$
the eigenvalues \citep{MandelWolf}.

Another possibility of introducing eigenmodes is the Schmidt decomposition.
Such a modal decomposition describes efficiently the correlations
in a bipartite system. In particular, it can be used to describe twin beams (idler and signal) at the output of high-gain PDC~\citep{Sharapova:15}. Although a deeper consideration shows that the shapes of the Schmidt modes change with the parametric gain~\citep{Sharapova:18}, this effect is small. Accordingly, one can obtain these modes from the decomposition of the so-called two-photon amplitude (TPA), i.e. the probability amplitude of signal/idler photons being emitted
into plane-wave monochromatic modes with transverse wave-vectors $\vec{q}_{s,i}$. Provided that the PDC generation has an axial symmetry (for instance, spatial walk-off is small), the TPA can be written as
\begin{equation}
F\left(\vec{q}_{s},\vec{q}_{i}\right)=\sum_{l,p}\sqrt{\lambda_{lp}}\frac{u_{lp}(\theta_{s})}{\sqrt{\theta_{s}}}e^{il\phi_{s}}\frac{v_{lp}(\theta_{i})}{\sqrt{\theta_{i}}}e^{-il\phi_{i}},\label{eq:TPA}
\end{equation}
where $l,p$ are, respectively the azimuthal and radial indices, $\theta=q/k$ the radial angle,  $k$ the wavevector modulus, and $\phi=\tan^{-1}q_{y}/q_{x}$ the azimuthal angle. The coefficients $\lambda_{lp}$ are the Schmidt eigenvalues with the normalization condition $\sum_{l,p}\lambda_{lp}=1$
and $u_{lp},v_{lp}$ are the orthonormal radial modes for signal and
idler, considered to be equivalent in the degenerate regime. We would like to point
out that if $l\ne0$, the azimuthal eigenmodes are periodic functions
of $\phi$ and a mode with number $l$ carries an OAM charge $l$.

One can show that the coherent modes (\ref{eq:cohmodes}) and eigenvalues of one of the twin beams (signal or idler) coincide with the Schmidt modes and Schmidt coefficients given in Eq.~(\ref{eq:TPA}),
i.e. $\tilde{\lambda}_{lp}={\lambda_{lp}}$ and $\tilde{u}_{lp}(\vec{q})=\frac{u_{lp}(\theta)}{\sqrt{\theta}}e^{il\phi}$
\citep{Bobrov:13,Felix}.  

Consider the covariance of intensities measured at angles $\theta,\phi$ and $\theta',\phi'$, defined as
\begin{equation}
{\rm Cov}(\theta,\theta',\phi,\phi')\equiv\left<I(\theta,\phi)I(\theta',\phi')\right>-\left<I(\theta,\phi)\right>\left<I(\theta',\phi')\right>.
\label{eq:covar}
\end{equation}
In terms of the Schmidt modes introduced in Eq.~(\ref{eq:TPA}), this covariance has the form~\citep{Finger:17,Beltran:17}
\begin{equation}
\begin{array}{c}
{\rm Cov}\left(\theta,\theta',\phi,\phi'\right)=\left[\sum_{l,p}\lambda_{lp}\frac{u_{lp}\left(\theta\right)}{\sqrt{\theta}}e^{il\phi}\frac{u_{lp}^{*}\left(\theta'\right)}{\sqrt{\theta'}}e^{-il\phi'}\right]^{2}\\
+\left[\sum_{l,p}\lambda_{lp}\frac{v_{lp}\left(\theta\right)}{\sqrt{\theta}}e^{il\phi}\frac{v_{lp}^{*}\left(\theta'\right)}{\sqrt{\theta'}}e^{-il\phi'}\right]^{2}
+2\left|\sum_{l,p}\lambda_{lp}\frac{u_{lp}\left(\theta\right)}{\sqrt{\theta}}e^{il\phi}\frac{v_{lp}\left(\theta'\right)}{\sqrt{\theta'}}e^{il\phi'}\right|^{2}.
\label{eq:cov}
\end{array}
\end{equation}

To simplify the reconstruction, in experiment we eliminate one of the twin beams by using frequency filtering that is asymmetric with respect to the degenerate wavelength. In this way, we transmit
the signal photons and remove their idler matches to study only the autocorrelations of the intensity fluctuations. Then the last two terms in Eq.~(\ref{eq:cov}) disappear, and the covariance takes the form
\begin{equation}
{\rm Cov}(\theta,\theta',\phi,\phi')=\left[\sum_{l,p}\lambda_{lp}\frac{u_{lp}(\theta)}{\sqrt{\theta}}e^{il\phi}\frac{u_{lp}^{*}(\theta')}{\sqrt{\theta'}}e^{-il\phi'}\right]^{2}.
\label{eq:autocorr}
\end{equation}
This relation allows the mode shapes to be reconstructed from the square root of the covariance using the singular value decomposition.

\section{Radiation eigenmodes at the output of the SU(1,1) interferometer\label{sec:SU(1,1)}}

In this Section, we focus our attention on the eigenmode decomposition
of the radiation at the output of the wide-field ${\rm SU(1,1)}$
interferometer \citep{Frascella:19}. 

\subsection{Theory}
We derive the Schmidt decomposition for this device using the following theoretical approach. The emission from first crystal is considered as the input for the second crystal, i.e. an optical parametric amplifier \citep{Perez:14}. We do not have to take into account the diffraction of the emission because of the focusing element. The parametric gain values in the two crystals are assumed to be different, $G_1\ne G_2$, and the phase of the SU(1,1) interferometer is $\Phi$. We describe the state evolution by using the squeezing operators $\hat S_1$, $\hat S_2$, respectively, for the first and second amplifier. The state at the output of the system is
\begin{equation}
\left|\psi \right> = \hat{S}_2 \hat{S}_1 \left| 0 \right>,
\end{equation} 
where $\left| 0 \right>$ is the vacuum state.

By using the decomposition of the TPA in Eq.~(\ref{eq:TPA}), one can define the photon annihilation operators for the collective spatial Schmidt modes \citep{Sharapova:15}. In our case, these Schmidt-mode operators at the output of the second crystal can be expressed in terms of the initial (vacuum) photon annihilation operators  $\hat A^{\rm{in}}_{lp} $ and $\hat B^{\rm{in}}_{lp} $ in the corresponding $(l,p)$ Schmidt modes as
\begin{equation}
\begin{array}{c}
\hat A^{\rm{out}}_{lp} = \hat{S}_1^{\dagger} \hat{S}_2^{\dagger}  \hat A^{\rm{in}}_{lp} \hat{S}_2 \hat{S}_1 = \hat{S}_1^{\dagger} (\hat A^{\rm{in}}_{lp} \cosh \sqrt{\lambda_{lp}} G_2 + \left[\hat B^{\rm{in}}_{lp} \right]^{\dagger} e^{i \Phi} \sinh \sqrt{\lambda_{lp}} G_2 )\hat{S}_1 = \\
= (\hat A^{\rm{in}}_{lp} \cosh \sqrt{\lambda_{lp}} G_1 + \left[ \hat B^{\rm{in}}_{lp} \right]^{\dagger} \sinh \sqrt{\lambda_{lp}}G_1) \cosh \sqrt{\lambda_{lp}} G_2 + \\+(\left[ \hat B^{\rm{in}}_{lp} \right]^{\dagger} \cosh \sqrt{\lambda_{lp}} G_1 + \hat A^{\rm{in}}_{lp} \sinh \sqrt{\lambda_{lp}}G_1) e^{i \Phi} \sinh \sqrt{\lambda_{lp}}G_2.
\end{array}
\label{Heiseqs}
\end{equation}

By defining the quantities
\begin{equation}
\begin{array}{c}
w^{\rm{eff}}_{1,lp}(G_1,G_2, \Phi) = \cosh \sqrt{\lambda_{lp}}G_1 \cosh \sqrt{\lambda_{lp}}G_2 + e^{i\Phi} \sinh \sqrt{\lambda_{lp}}G_1 \sinh \sqrt{\lambda_{lp}}G_2,
\\
w^{\rm{eff}}_{2,lp}(G_1,G_2, \Phi) = \sinh \sqrt{\lambda_{lp}}G_1 \cosh \sqrt{\lambda_{lp}}G_2 + e^{i\Phi} \cosh \sqrt{\lambda_{lp}}G_1 \sinh \sqrt{\lambda_{lp}}G_2,
\end{array}
\end{equation}
Eq.~(\ref{Heiseqs}) can be simplified to
\begin{equation}
\hat A^{\rm{out}}_{lp} =w^{\rm{eff}}_{1,lp}(G_1,G_2, \Phi) \hat A^{\rm{in}}_{lp} + w^{\rm{eff}}_{2,lp}(G_1,G_2, \Phi) \left[ \hat B^{\rm{in}}_{lp} \right]^{\dagger}.
\end{equation}

With this result, it is possible to calculate the average number of signal photons in the plane-wave mode $\vec q$~\citep{Sharapova:15},
\begin{equation}
\begin{array}{c}
\left<\hat N_s(\vec q)\right> = \sum_{l,p} \left| w^{\rm{eff}}_{2,lp}(G_1,G_2, \Phi) \right|^2 \frac{\left|u_{lp}(\theta)\right|^2}{\theta}.
\end{array}
\end{equation}
By integrating over the radial angle $\theta$, we get the average total number {of signal photons,}
\begin{equation}
\left< \hat N^{\rm{tot}}_s \right>= \sum_{l,p} \left| w^{\rm{eff}}_{2,lp}(G_1,G_2, \Phi) \right|^2,
\end{equation}
as a sum of the terms $\left| w^{\rm{eff}}_{2,lp}(G_1,G_2, \Phi) \right|^2$ that we identify as the number of photons in the $(l,p)$ Schmidt mode.
In this fashion, we can denote the signal Schmidt eigenvalues as
\begin{equation}
\Lambda_{lp} = \left| w^{\rm{eff}}_{2,lp}(G_1,G_2, \Phi) \right|^2.
\end{equation}

If the gains are large enough to satisfy the relations $\sqrt{\lambda_{lp}}G_{1/2}\gg1$ and provided that $\Phi$ is not close to $\pi$, we can approximate
\begin{equation}
\Lambda_{lp}\approx \frac{1}{4} \sinh^2 ( \sqrt{\lambda_{lp}}(G_2+G_1) ) \left|1+e^{i \Phi}\right|^2.
\label{approxweights}
\end{equation}
This approximation is valid as long as 
\begin{equation}
\sinh\left(\sqrt{\lambda_{lp}}\left(G_1+G_2\right)\right)\left|1+e^{i \Phi}\right|\gg\sinh\left(\sqrt{\lambda_{lp}}\left|G_1-G_2\right|\right),
\end{equation}
hence the condition of $\Phi\ne\pi$.

\subsection{Experimental setup}

\begin{figure}[H]
\begin{centering}
\includegraphics[width=0.6\textwidth]{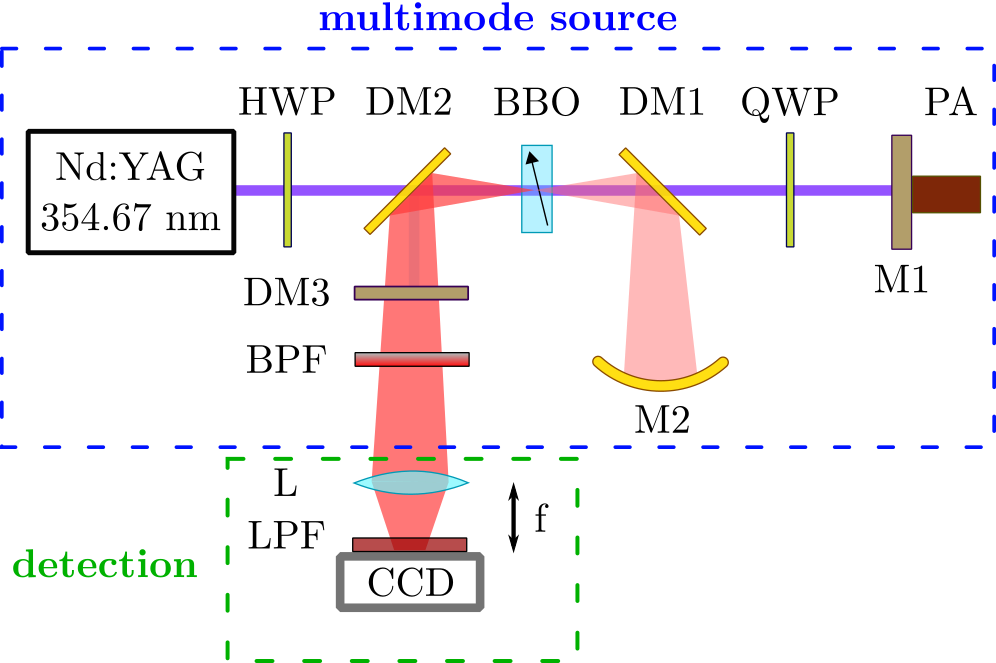}
\par\end{centering}
\caption{Experimental setup.}
\label{fig:setup}
\end{figure}

The experimental setup is shown in Fig. \ref{fig:setup}. The basic parts of the interferometer are a $\beta$-barium borate (BBO) crystal and mirrors M1 and M2. The pump is a Nd:YAG laser
emitting $18$ ps pulses at the wavelength $354.67$ nm with a repetition
rate of $1$ kHz. We produce type-I PDC radiation in the collinear
degenerate regime at $709.33$ nm. The dichroic mirror DM1 separates the pump and the down-converted radiation. The pump is reflected back to the crystal by
the mirror M1 mounted on the piezoelectric actuator PA in order to
control the phase of the interferometer. The PDC radiation is imaged
by the spherical mirror M2 ($R=100$ mm) back onto the crystal.
With an average pump power of $60$ mW and a FWHM of $300\pm10$ ${\rm \mu}m$,
the PDC emission in the first pass of the pump beam is multimode even
if the parametric gain is high, i.e. $G_{1}=2.1\pm0.3$ (measured
as in Ref. \citep{Spasibko:12}). We select a higher gain in the second
pass, $G_{2}=3.3\pm0.3$, by focusing the pump beam to a FWHM of $180\pm10$
${\rm \mu}m$ and controlling the polarisation with the waveplates
HWP and QWP. Indeed, the polarisation is rotated from horizontal to
27 deg by the half-wave plate HWP in order to reduce the contribution
in the first pass and rotated back to horizontal with the double pass
through the quarter wave-plate QWP.

In the second pass, if the pulses meet again at the crystal, the multimode
structure undergoes amplification/de-amplification depending on
the relative phase between pump, signal and idler beams $\Phi=\Phi_{p}-\Phi_{s}-\Phi_{i}$.
The resulting multimode radiation is very similar to the one of a
single crystal with the effective gain $G_{1}+G_{2}$, as shown in Eq.~\ref{approxweights}.
The dichroic mirror DM2 reflects the amplified PDC emission to the
detection part, while the dichroic mirror DM3 rejects any residual
pump beam. The band-pass filter BPF (central wavelength $700$ nm, bandwidth
$10$ nm) selects a frequency band slightly shifted from the degeneracy, so that only one of the twin beams is detected. At the same time, the shift from degeneracy is so small that we do not expect the signal and idler eigenmodes to be different.  A
charge-coupled device (CCD) camera is placed in the far field of the
lens L (focal length $f=4$ cm) and has a long-pass filter LPF attached
to reduce the contribution of the stray light to the intensity
distributions.

\begin{figure}[H]
\begin{centering}
\includegraphics[width=0.55\textwidth]{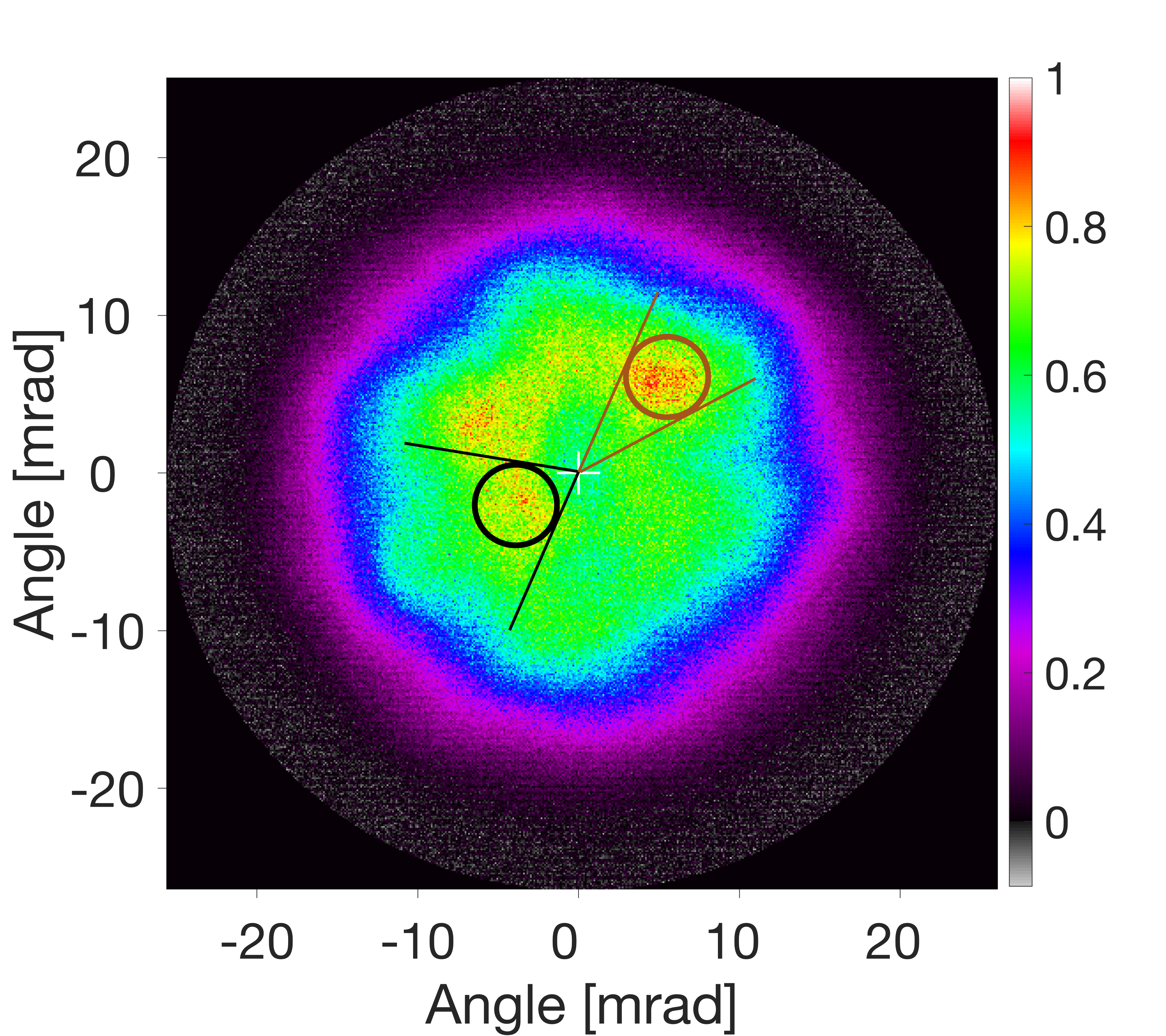}
\par\end{centering}
\caption{Single-shot far-field 2D intensity distribution at the output of the
wide-field ${\rm SU(1,1)}$ interferometer. Two speckles of intensity
fluctuations are shown by circles. The phase of the
interferometer $\Phi$ is $3.82$ rad.}
\label{fig:speckles}
\end{figure}

Figure \ref{fig:speckles} shows the far-field intensity distribution recorded in a single pulse. The distribution is over two Cartesian angles defined
as $q_{x}/k$ and $q_{y}/k$ and it has a FWHM of $\sim22$ mrad.
In this case, the phase $\Phi$ is fixed to the value $3.82$ rad,
i.e. approximately $\pi/5$ away from the dark fringe. This is achieved
with a phase-locking arrangement not shown for simplicity in the setup
in Fig. \ref{fig:setup}. Typically, the spectrum of high-gain PDC radiation would contain correlated intensity fluctuations at points symmetric with respect to the pump direction (center). Here, these correlations are not present
because only one of the twin beams is selected by non-degenerate filtering. The black and brown circles highlight especially bright areas (`speckles') of the PDC
emission \citep{Brida:09}. 

\subsection{Orbital angular momentum spectrum}

For the measurement of the OAM spectrum at different radial angles,
we keep the phase $\Phi=3.82$ rad. We acquire $500$ single-shot
2D intensity distributions and extract the azimuthal intensity
1D distribution $I(\phi)$ by integrating $I(\theta,\phi)$
over a range of $\pm1.1$ mrad around each radial angle $\theta_{0}$.
For the resulting 1D distributions, the covariance for the intensities
$I(\phi)$, $I(\phi')$ at the azimuthal angles
$\phi$ and $\phi'$ is experimentally obtained as

\begin{equation}
\mathrm{Cov}_{\left|\theta=\theta'=\theta_{0}\right.}(\phi,\phi')=\left<I(\phi)I(\phi')\right>-\left<I(\phi)\right>\left<I(\phi')\right>.
\end{equation}
Then, the autocorrelation part in Eq.~(\ref{eq:autocorr}) takes the
form
\begin{equation}
\mathrm{Cov}_{\left|\theta=\theta'=\theta_{0}\right.}(\phi,\phi')=\left[\sum_{l}\mathcal{L}_{l}(\theta_{0})e^{il(\phi-\phi')}\right]^{2},\label{eq:azimcov}
\end{equation}
with $\mathcal{L}_{l}(\theta_{0})=\sum_{p}\Lambda_{lp}\left|u_{lp}(\theta_{0})\right|^{2}/\theta_{0}$,
taking into account the high-gain weights given by Eq.~(\ref{approxweights}).
The covariance distribution {turns out to be symmetric and depends only on} the difference $\phi-\phi'$.

\begin{figure}[H]
\begin{centering}
\includegraphics[width=0.75\textwidth]{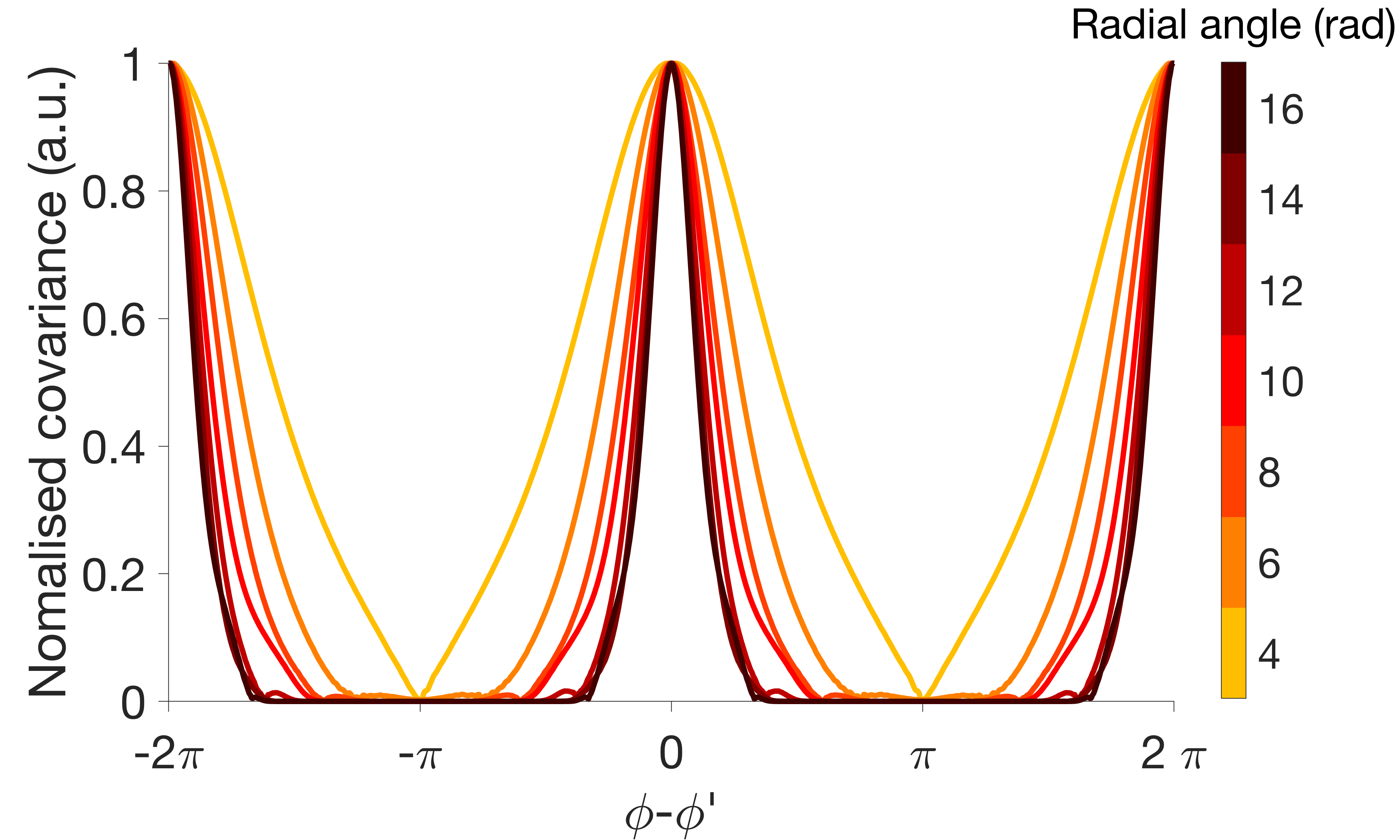}
\par\end{centering}
\caption{Experimental 1D azimuthal covariance distributions as functions of
the difference in the azimuthal angles $\phi-\phi'$ with fixed radial
angles ranging from 4 to 16 mrad. The width of the distribution is
larger for smaller radial angles.}
\label{fig:azimcov}
\end{figure}

In experiment, we average the distribution along the $\phi+\phi'$
direction and the result of this procedure is shown in Fig.~\ref{fig:azimcov}.
Since the possible range for the azimuthal angle is between 0 and
$2\pi$, the difference $\phi-\phi'$ spans between $-2\pi$ and $2\pi$.
For $\phi=\phi'$, the
intensity fluctuations are maximally correlated, as expected for the
autocorrelation. At different azimuthal angles $\phi\ne\phi'$, the correlation drops to zero. 

If we fix the radial angle $\theta_0$ to be small, i.e. choose points
closer to the center, the covariance distribution is broader. Conversely,
it gets tighter for large {$\theta_0$}.

\begin{figure}[H]
\begin{centering}
\includegraphics[width=0.6\textwidth]{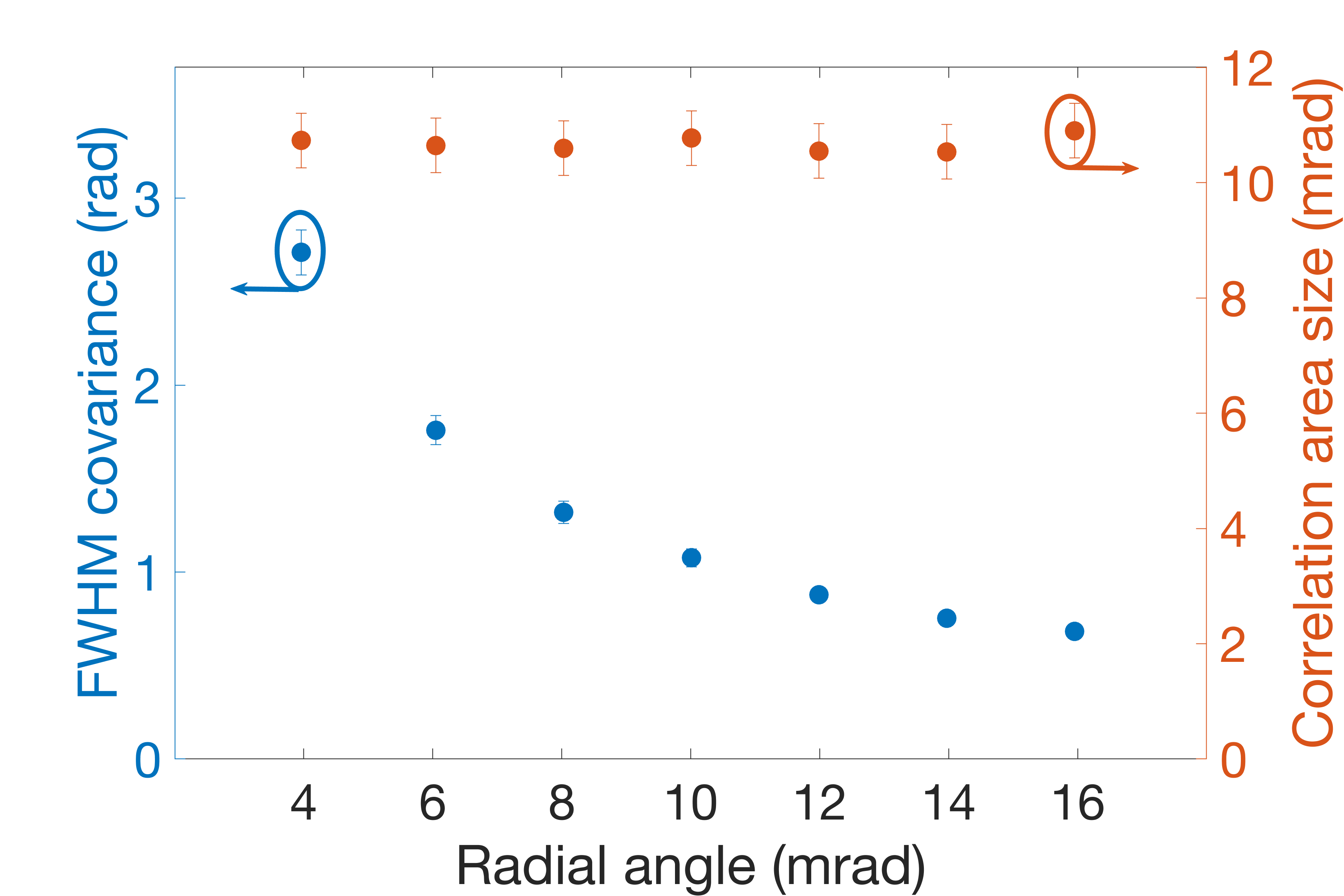}
\par\end{centering}
\caption{FWHM of the azimuthal covariance distribution shows a decrease as
a function of the radial angle. By multiplying this width by each
radial angle, we see that the speckle size in Cartesian angles
is constant.}
\label{fig:azimFWHM}
\end{figure}

This trend is presented on the left y-axis in blue of Fig. \ref{fig:azimFWHM}.
The covariance has a FWHM ranging from $2.7$ to 0.7 mrad respectively
for radial angles $4$ and $16$ mrad. As one can see in Fig. \ref{fig:speckles},
the azimuthal size of the correlation area~\citep{Kumar:18,Brida:09} is larger for smaller radial angles. At the same time, the correlation area size should be the same in Cartesian angles $q_x/k,\,q_y/k$ as the speckle size in Fig.~\ref{fig:speckles} is also the same everywhere.
By multiplying the covariance FWHM by the selected radial angle, we see that 
the correlation area size is constant as expected. We plot the result in orange on
the right y-axis of Fig. \ref{fig:azimFWHM} and we find an average
value of $11\pm1$ mrad. 
\begin{figure}[H]
\begin{centering}
\includegraphics[width=0.65\textwidth]{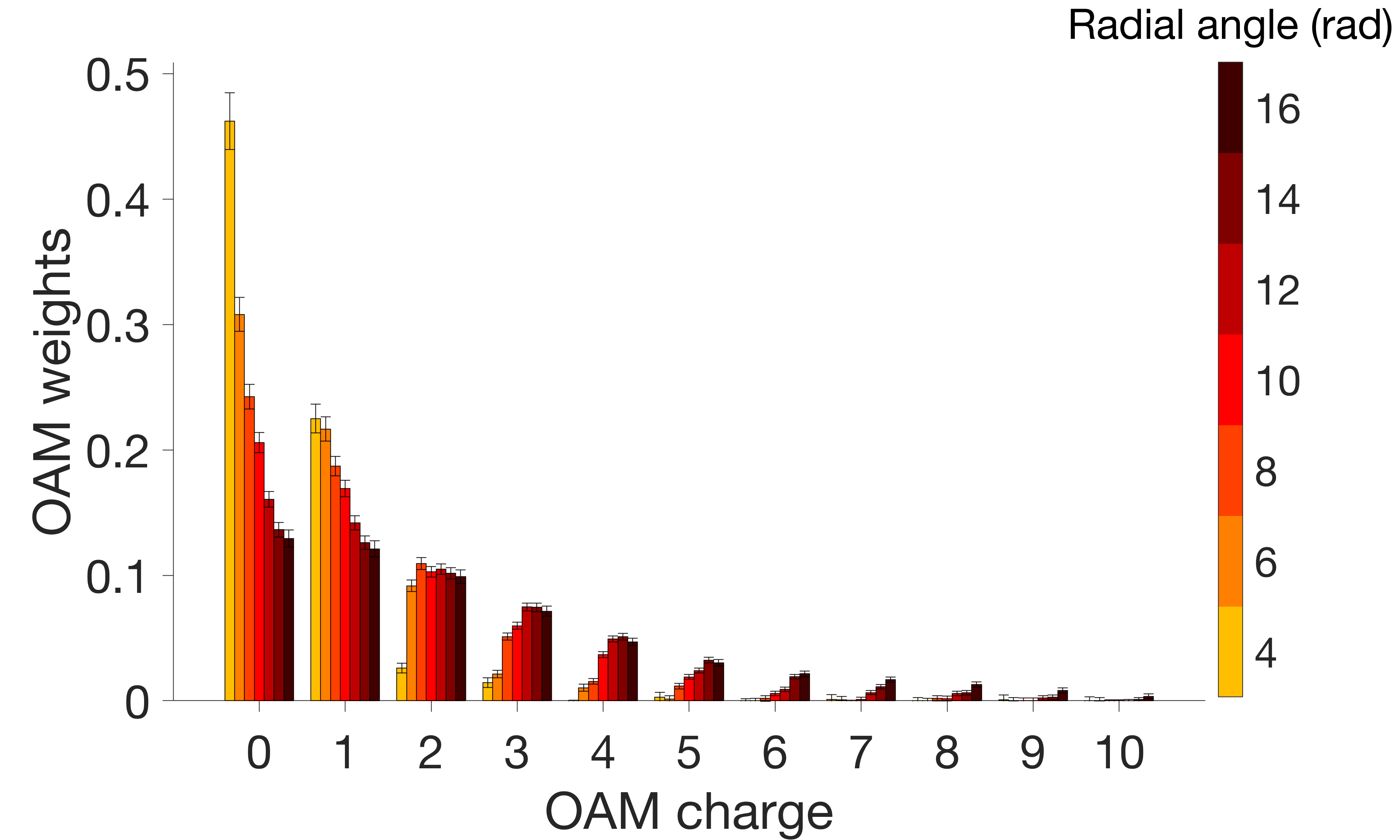}
\par\end{centering}
\caption{OAM spectrum obtained from the covariance measured at radial angles
ranging from $4$ to $16$ mrad. The distributions show a larger number
of azimuthal modes for larger radial angles.}
\label{fig:OAMrad}
\end{figure}

By applying the Fourier decomposition to the square root of the covariance~(\ref{eq:azimcov}),
we obtain the distribution of weights $\mathcal{L}_{l}(\theta_{0})$
for the radial angle considered, as shown in Fig.~\ref{fig:OAMrad}.
The part of negative OAM charge is not shown because it is symmetric with
respect to $l=0$. For larger radial angles, the covariance has a steeper dependence on
$\phi-\phi'$, which corresponds to a larger number of OAM modes~\citep{Fedorov:16}. Accordingly, the
number of modes, calculated as $1/\left[\mathcal{L}_{l}(\theta_{0})\right]^{2}$,
increases from $3.8\pm0.2$ to $18.2\pm0.6$ for $\theta_0$ increasing from
$4$ mrad to $14$ mrad.

The weights
$\mathcal{L}_{l}(\theta_{0})$ obtained from the covariance at different $\theta_{0}$ contain contributions from
several radial modes. Therefore, we cannot determine
the distribution of the weights $\Lambda_{lp}$ for a fixed $p$.
Nevertheless, it is possible to obtain the average OAM weights over
all radial modes, i.e. $\Lambda_{l}=\sum_{p}\Lambda_{lp}$, by integrating
the square root of the covariance~(\ref{eq:azimcov}) over the radial angle,
\begin{equation}
\int\sqrt{\mathrm{Cov}_{\left|\theta=\theta'=\theta_{0}\right.}(\phi,\phi')}\theta_{0}d\theta_{0}=\sum_{l}\Lambda_{l}e^{il(\phi-\phi')}.\label{eq:azimcovavg}
\end{equation}
This procedure has been performed in Ref.~\citep{Frascella:19} and resulted in $8$ OAM modes.

It is interesting to note that according to Eq.~(\ref{approxweights}), the OAM spectrum remains the same for different phases of the interferometer. This result has been experimentally confirmed in Ref. \citep{Frascella:19}.

\subsection{Radial modes}

For reconstructing the radial modes, we integrate the measured distribution $I(\theta,\phi)$ within
a range of $\pm0.08$ rad around $\phi=0$ to obtain the radial distribution
of intensity $I(\theta)$. The radial covariance distribution is then obtained as

\begin{equation}
\mathrm{Cov}_{\left|\phi=\phi'=0\right.}(\theta,\theta')=\left<I(\theta)I(\theta')\right>-\left<I(\theta)\right>\left<I(\theta')\right>.
\end{equation}

\begin{figure}[H]
\begin{centering}
\includegraphics[width=0.4\textwidth]{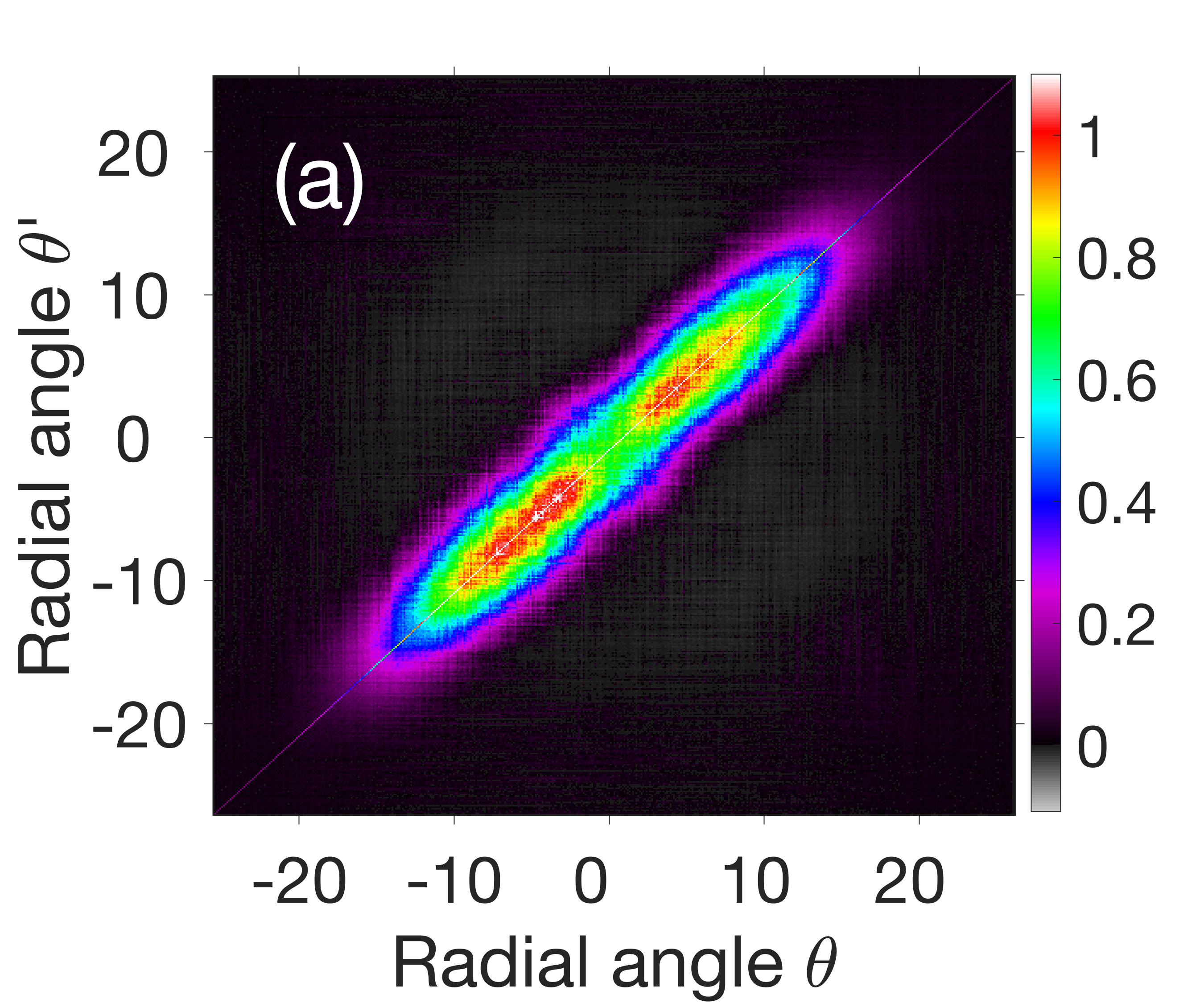}\includegraphics[width=0.4\textwidth]{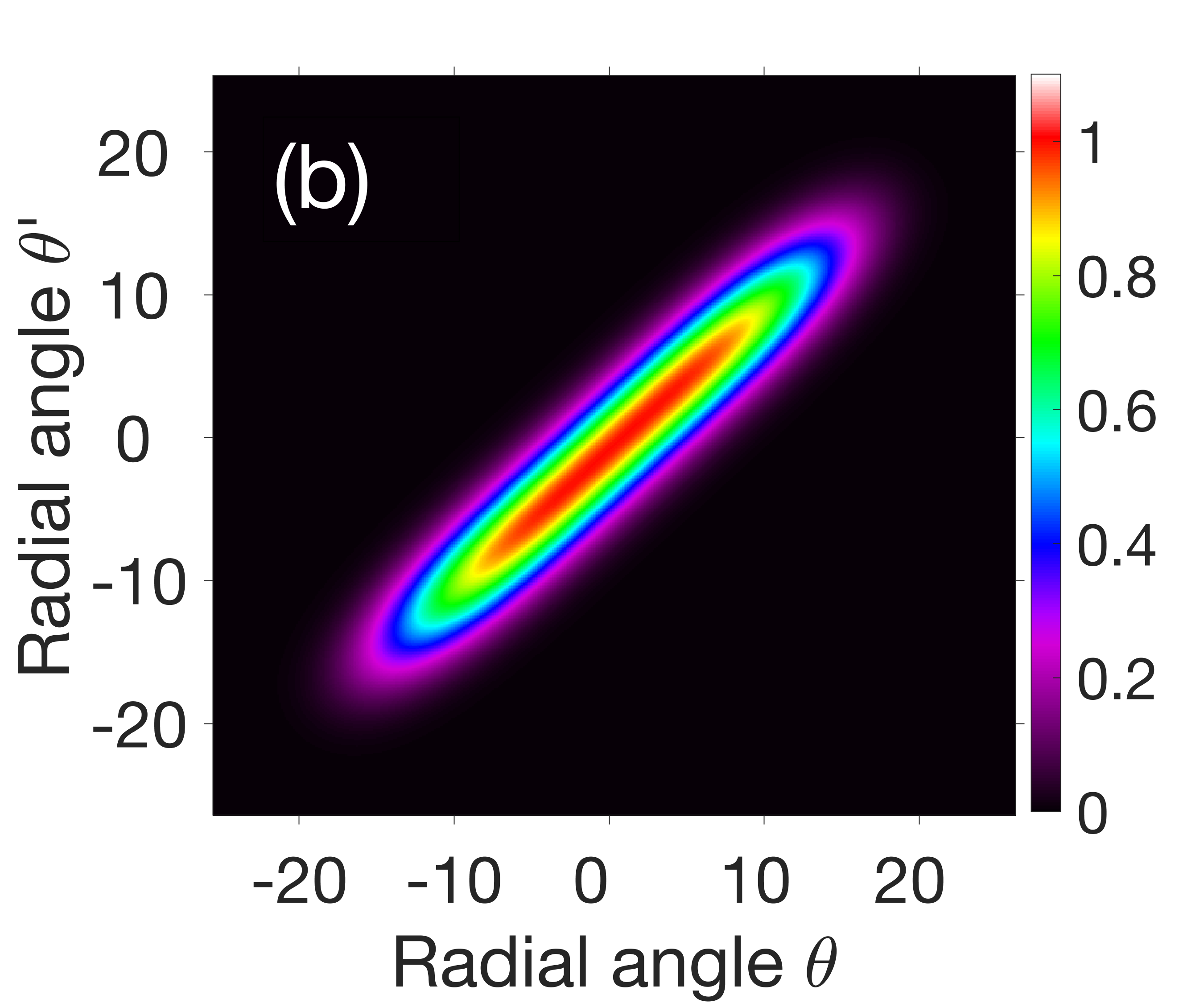}
\par\end{centering}
\caption{(a) Covariance distribution for the radial degree of freedom with $\Phi=3.82$ rad. The
correlation {is distributed around the $\theta=\theta'$ direction.} (b) Theoretical fit of the distribution.}
\label{fig:radcovar}
\end{figure}

The results for $\Phi=3.82$ rad are shown in Fig. \ref{fig:radcovar} (a). The correlation of intensity fluctuations is the highest along the diagonal $\theta=\theta'$. Somewhat lower values at small radial angles are probably caused by a damaged spot on the optical elements. This cannot be usually recognized in the single-shot spectrum, but it becomes clear by averaging over several pulses. For this reason, we use a fitting function for the covariance as described in Ref.~\citep{Sharapova:18}. The fit requires prior knowledge about the shape of the intensity distribution and the nature of the correlations, in general not necessary for the reconstruction of the modes. The result of the fit is shown in Fig. \ref{fig:radcovar} (b) and it shows good agreement with the experimental covariance.

In this case, the covariance in Eq.~(\ref{eq:autocorr}) takes the form
\begin{equation}
\mathrm{Cov}_{\left|\phi=\phi'=0\right.}(\theta,\theta')=\left[\sum_{p}\Lambda_{p}\frac{u_{p}(\theta)}{\sqrt{\theta}}\frac{u_{p}^{*}(\theta')}{\sqrt{\theta'}}\right]^{2},\label{eq:radcov}
\end{equation}
with $\Lambda_{p}=\sum_{l}\Lambda_{lp}$ and assuming that the radial
modes do not depend strongly on $l$~\citep{Beltran:17}. By performing the singular value
decomposition on the square root of Eq.~(\ref{eq:radcov}), we obtain
the shapes and the weights of the Schmidt radial modes.

\begin{figure}[H]
\begin{centering}
\includegraphics[width=0.45\textwidth]{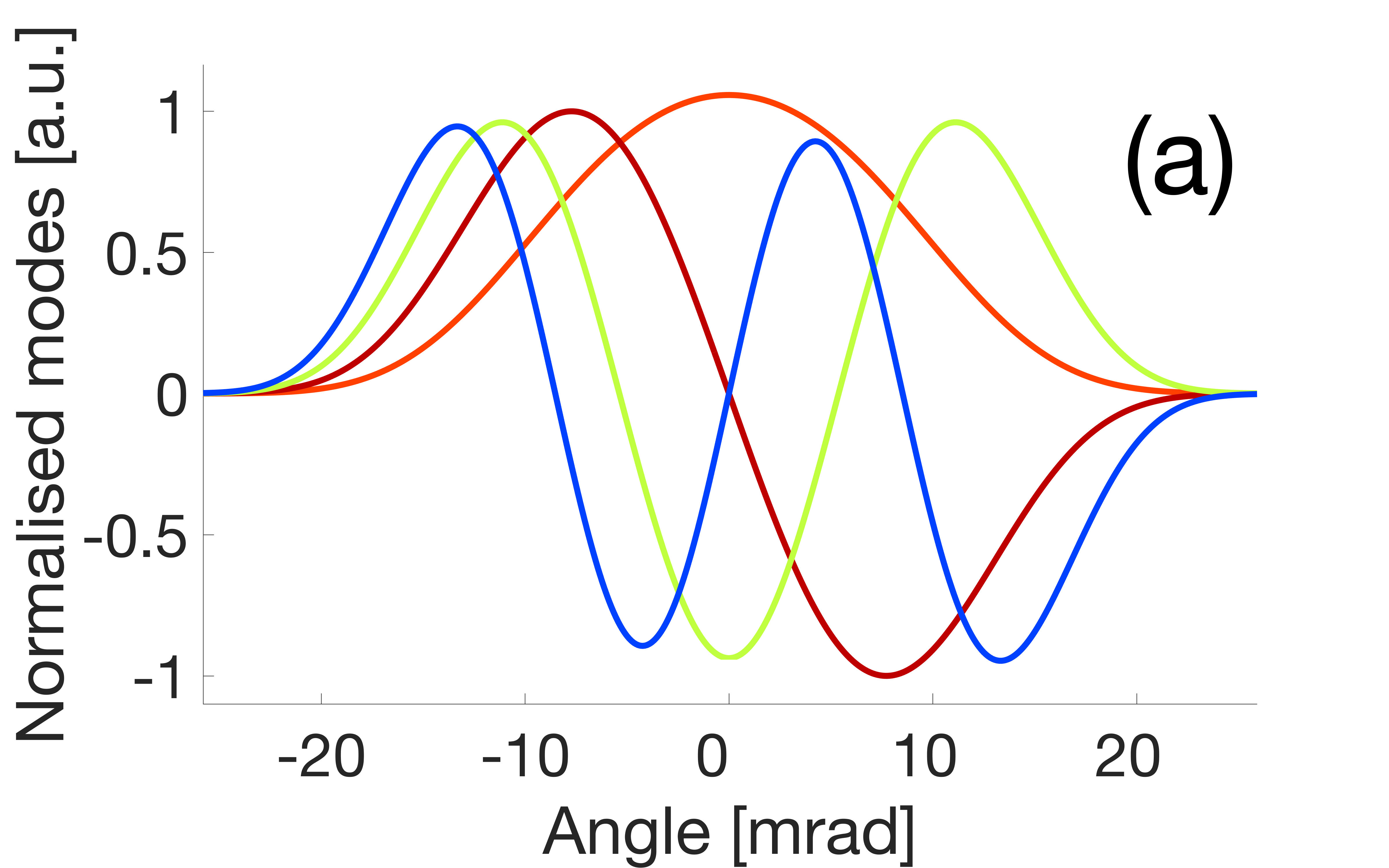}\includegraphics[width=0.46\textwidth]{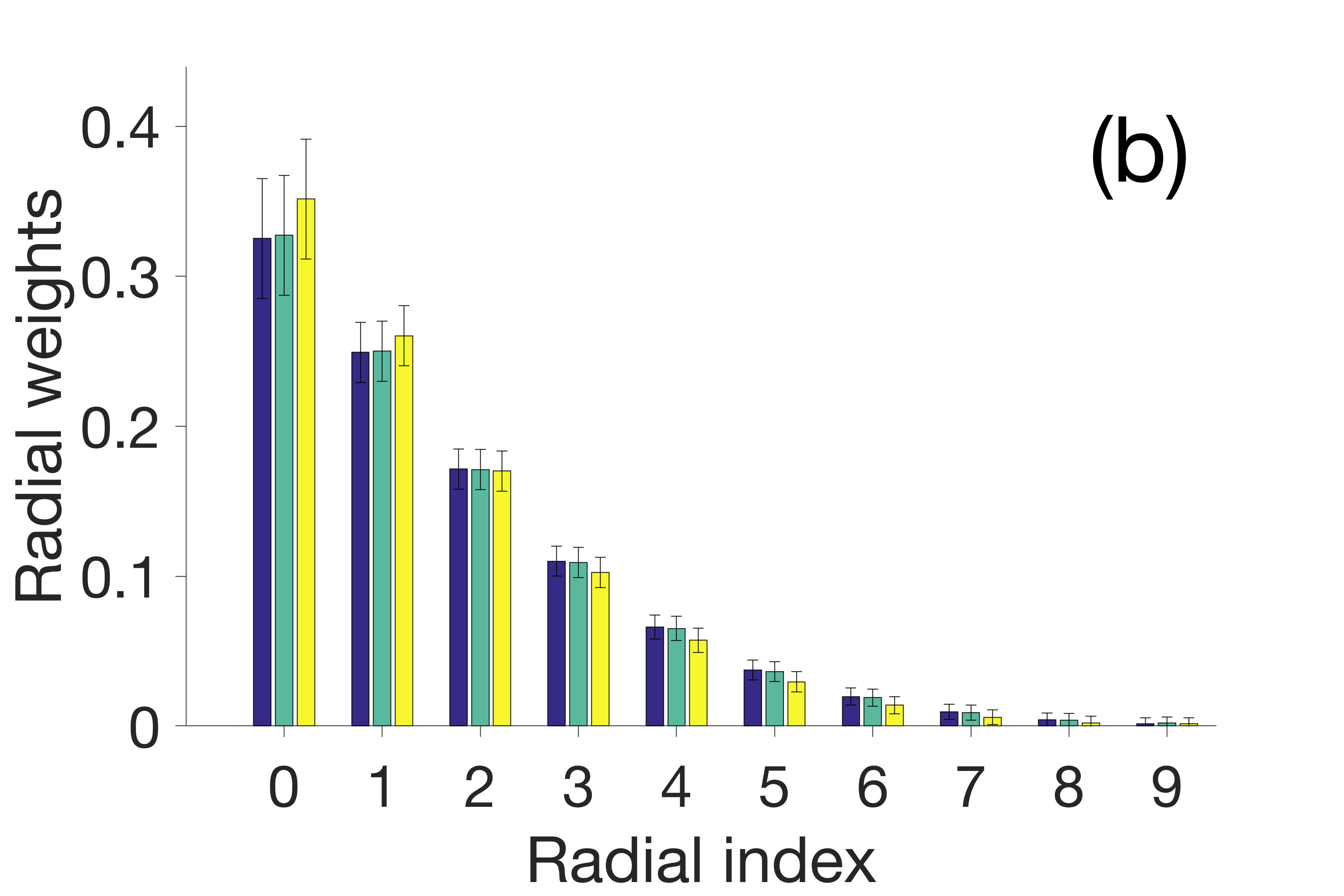}
\par\end{centering}
\caption{Shapes (a) and weights (b) of the radial modes. Only the first four
mode shapes are shown (first, second, third and fourth in the colors
orange, red, green and blue). The weights are shown for three different
phases of the interferometer, namely $\Phi=3.82$, $4.02$ and $4.22$
rad correspondingly to the colors blue, green and yellow.}
\label{fig:radialmodes}
\end{figure}

The result is shown in Fig. \ref{fig:radialmodes}. The shapes of the first four modes resemble the Hermite-Gauss ones, as one would expect for a covariance distribution close to a Gaussian shape.
The distribution of the weights is constant as the phase of the interferometer
is changed between the value $\Phi=3.82$, $4.02$ and $4.22$ rad
(color blue, green and yellow). The number of modes in this case is $4.5\pm0.2$. Accordingly, the total number of spatial modes can be estimated as $34\pm2$, which roughly corresponds to the number of speckles within the intensity distribution at the output of the interferometer (Fig.~\ref{fig:speckles}).

\section{Conclusion}

We have demonstrated the experimental reconstruction of the eigenmodes
of a multimode field with Gaussian intensity fluctuations by using the measurement of
intensity distributions. To this end, we have justified theoretically the link between
the measurement of intensity fluctuations correlation and the equivalent
coherent-mode or Schmidt decomposition in the cases of thermal light and bright twin beams.
For the experiment, we employed a wide-field ${\rm SU(1,1)}$ interferometer
to generate twin beams and used one of the two beams
as a state of light with thermal statistics. By acquiring $500$ single-shot far-field
intensity spectra we reconstructed the OAM spectrum of our source
for several radial angles and the spectrum averaged over the radial
degree of freedom. We also determined the shapes and weights of the
radial modes.

The method we portrayed in our work has the advantage of a simple
detection scheme, requiring only a camera measuring intensity distributions.
Moreover, no prior knowledge about the modes of the system under study
is needed, but simple considerations on the symmetry of the intensity
distribution simplify the treatment. Since the reconstruction relies
on the acquisition of several intensity spectra, instant changes of
the mode composition cannot be detected. Another critical point is
that the detection of intensity distributions leaves out the information
about the space-dependent phases of the modes.

This technique should be suitable to quickly identify
the mode structure of an unknown stable system. For instance, we believe
one could measure the eigenmode decomposition of a fiber whose guiding
properties are completely unknown. By measuring at the output the intensity
distributions of thermal radiation launched into the fiber, the shapes
of the modes can be obtained easily, while the relative strengths
might depend on the coupling of the source. Currently, the most widely
used method is the $S^{2}$ imaging, which relies on the group delay
difference between different modes propagating and interfering in
the fiber \citep{Nicholson:08}. For such a method, a tunable source
is essential to provide the mode shapes and spectrum \citep{Nguyen:12},
therefore our reconstruction could offer a valid alternative.

For the characterization of an SU(1,1) interferometer, an important conclusion is that the use of a focusing element provides a multimode spatial spectrum at the output: in total, more than $30$ spatial modes were measured. This is not the case in a configuration where such an element is absent~\citep{Beltran:17}. Moreover, the output number of OAM modes can be large (up to $18$) if a large radial angle is selected. Although low-gain PDC provides even larger numbers of OAM modes~\citep{Kulkarni:17}, at high gain this number is usually reduced. The possibility to have an SU(1,1) interferometer covering a large number of OAM modes paves the way to sub-shot-noise OAM sensing.

\ack{}{}

We acknowledge financial support of the Russian Science Foundation project No.19-42-04105.
We would like to thank Valentin A. Averchenko for the fruitful discussions.

\bibliographystyle{unsrt}
\addcontentsline{toc}{section}{\refname}\bibliography{spatcorr}

\end{document}